\documentclass[twocolumn]{paper}
\usepackage{amsmath,amssymb}
\usepackage{graphicx}
\usepackage[english]{babel}

\author{Lennart Hilbert\thanks{Department of Physiology, McGill University, Center for Applied Mathematics in Biosciences and Medicine and Meakins-Christie Laboratories, McGill University; McIntyre Medical Building, 3655 Promenade Sir William Osler, H3G 1Y6 Montreal, Quebec, Canada, Tel. +1 514 398-8092, Fax +1 514 398-7452, Lennart.Hilbert@mail.mcgill.ca}}

\title{Shifting gears: Thermodynamics of genetic information storage suggest stress-dependence of mutation rate, which can accelerate adaptation.}

\date{\today}

\begin{document}

\maketitle

\textbf{Running title:} Thermodynamics of genetic information storage

\begin{abstract}
\textit{Background:} Acceleration of adaptation dynamics by stress-induced hypermutation has been found experimentally. Evolved evolvability is a prominent explanation. We investigate a more generally applicable explanation by a physical constraint.

\textit{Methods and Results:} A generic thermodynamical analysis of genetic information storage obviates physical constraints on the integrity of genetic information. The capability to employ metabolic resources is found as a major determinant of mutation probability in stored genetic information. Incorporation into a non-recombinant, asexual adaptation toy model predicts cases of markedly accelerated adaptation, driven by a transient increase of mutation rate. No change in the mutation rate as a genetic trait is required. The mutation rate of one and the same genotype varies dependent on stress level.

\textit{Implications:} Stress-dependent mutation rates are physically necessary and challenge a condition-independent genotype to mutation rate mapping. This holds implications for evolutionary theory and pathogen and cancer evolution.
\end{abstract}


\section*{Introduction}

\textbf{Rationale:} Conceptually, life can be understood as an informed, open organization of non-equilibrium thermodynamical processes, as the first chapters of most biochemistry textbooks will yield. This is a potent concept to understand the informational and energetic dealings of an organism, be it metabolism \cite{ChemicalCycleKinetics}, motility \cite{HillMuscleContraction}, the origin of life and early evolution \cite{Wicken,ArticleWicken,EigenRundumschlag,EarlyReplicons,EigenHyperCycle}, evolution in general \cite{PrigogineNicolis}, or pathogen evolution \cite{AntibioticsVariabilityReview,PolymeraseFidelities,SmallRNAMoleculesCompetition} to name but a few applications.

We investigate the thermodynamical underpinnings of storage of genetic information. In the light of adaptation dynamics, this analysis indicates that the adaptation of simple organisms can well be accelerated by a transient phase of stress-induced hypermutation. Stress-induced hypermutation is commonly understood as an organismic response that developed as a consequence of a specific selection scenario. While this is a credible explanation, our analysis obviates a generally applicable physical constraint on genetic information storage as an explanation of stress-induced hypermutation.

\textbf{Background -- Genetic integrity and evolutionary adaptation:}
Mutations are a potentially lethal threat to hereditary information. Consequently, elaborate mechanisms to counteract mutations and their effects have evolved. However, evolution of these very mechanisms and life in general requires mutations as a source of genetic variation. Let this fundamental ambivalence guide our brief review of the evolutionary role of mutator genotypes and stress-induced hypermutation.

In both cases organisms have been observed to exhibit an elevated mutation rate when faced with an adaptation challenge. This suggests an organismic ability to respond to selection pressure with increased mutation rates. Classically mutation and selection are understood as two entirely independent elements of the evolutionary adaptation process. Thus, the experimental findings seem to have bearing on the foundations of our understanding of evolutionary theory. This is the main similarity between mutator genotypes and stress-induced hypermutation. We will spend the rest of the introduction to clarify the fundamental differences of both concepts. We hope this prevents associations of stress-induced hypermutation with mutator genotype studies, which we found to frequently occur.

\textbf{Background -- Selection of mutator genotypes:}
Elevated mutation rates can be found by comparison between species, strains and even cells of the same population. A vast body of literature has discussed, and in our opinion mostly explained this finding. For a general review see \cite{EvolutionOfMutationRatesReview}. Mutator genotypes arise in response to specific selection scenarios that favor higher mutation rates \cite{MutationRateVariationMulticellular, Oenococcus, EvolutionOfMutationRatesExperimental}. High mutation rate genotypes outplay low mutation rate genotypes in scenarios of frequently changing selection criteria \cite{EvolutionOfMutationRates} and intricate mechanisms for efficient and effective evolvability have evolved \cite{EvolutionOfEvolvability, EvolutionOfEvolvabilityTwo}. Most importantly, elevated mutation rates of mutator genotypes are a selectable genetic trait. In this case the mutation rate is determined to a specific value from the genome of an organism, and this precise value is the result of evolutionary adaptation to specific environmental conditions.

\textbf{Background -- Stress-induced hypermutation:}
Cairn's Adaptive Mutation experiment \cite{OriginOfMutants} and several following studies have shown that genetic mutation rates in organisms of the exact same genotype can increase transiently under stress. This leads to accelerated adaptation to the present stress, which is followed by reduction of mutation rates to pre-stress level. While it is known that a change in selection criteria can lead to increased variation by broader spread of the population frequencies into existent genotypes, these findings suggests that altered selection criteria transiently increase the rate of occurrence of new genotypes. This questions the fundamental notion of independence of mutation rate from selective pressure as shown in the classic Luria-Delbrück experiment \cite{LuriaDelbrueck}. Quite the opposite, the common explanation of the Cairn's experiment is that evolution itself has awarded organisms with mechanisms to master evolution more efficiently and effectively \cite{EvolutionOfEvolvabilityTwo, OriginOfMutants, AdaptiveMutationIntroduction, MutationStressResponse, AmplificationMutationSeparate, StressResponseRegulation, StressMutationsFoster, StressMutationsEColi}.
The observed mutation rate increase is not caused by a genetic difference, but rather a dynamic organismic response to externally applied stress. Genetic alterations in cellular functions in general can also impose a stress, which in turn induces a secondary hypermutation response. Note that this response is not directly mediated by a genetic alteration in genes directly associated with mutation rate.

\section*{Methods}

\subsection*{Thermodynamical analysis of mutation suppression}

\textbf{General approach:} The genome stores an organism's hereditary information, and is therefore conceptually not different from any other information storage system: (1) The information storage system consists of a material medium that can take on different configurations. A specific configurations corresponds to a specific information stored in it. In DNA this is a specific sequence of base pairs. (2) Only rare and very specific configuration serve to store information in a useful way. For example, two DNA strands of the same length that contain the same information must at each position have the exact same bases. (3) These specific configurations are low entropy non-equilibrium states. Randomizing influences drive the system towards equilibrium, which corresponds to full loss of information. Only free energy expenditure can maintain the storage system close to its original configuration/information. DNA is constantly exposed to mutagenic (randomizing) influences, and only metabolically expensive repair mechanisms can maintain the original configuration. A deviation from the original configuration corresponds to genetic mutation. More generally speaking, the coupling of information storage and the second law is an example of the generalized Carnot principle \cite{BrillouinInformationAndScience}.

Common sense might already suggest that both an increase of the mutagenic influence or an inability to provide or utilize metabolic resources for mechanisms of mutation suppression/repair will lead to stronger mutagenic alteration of genetic information. Still, in the following we will construct a formal thermodynamic toy model to outline the optimum mutation suppression capability given a certain level of mutagenic influence and a certain level of metabolic resource (free energy) utilized for safe-guarding of the genetic information. The analysis is based on the definitions of a non-equilibrium thermodynamics description of Master equations as presented in \cite{NonequilibriumThermodynamics}.

\subsection*{Thermodynamical analysis}

Let a binary sequence $(b_n)\in\{0,1\}$ with $n=1,\dots ,N$ digits represent the binary genetic information of a generic organism. $0$ stands for a digit in its original value, while $1$ represents a digit whose value has been changed by mutation.

First, we will look at how the primary mutagenic influence and the metabolic resources utilized for the suppression of and protection against mutations set the probability $p$ of a single digit to contain its original value, a state represented by the value $0$. The genetic information storage system is a compartment at temperature $T$, which is in contact with a heat bath at temperature $T_{ext}$. $T_{ext}$ represents the strength of the primary mutagenic influence. Omnipresent mutagenic influences like molecular thermal collisions, background radiation or effects from reaction with surrounding chemical species warrant for $T_{ext}>0$ at all times under all conditions. Heat flows into the information storage system by diffusion and is exported by the use of metabolic resources:

\begin{align}
\frac{dQ}{dt} = c(T_{ext}-T) - T \cdot S^\prime_c,
\end{align}

where $c$ is a coefficient characteristic for the diffusion of heat from the heat bath into the information storage compartment. $S^\prime_c=(dS/dt)_c$ is the entropy production arising from the utilization of metabolic resource for the export of heat from the information storage system. For constant $S^\prime_c$ a globally stable steady state will be attained:

\begin{align}
dQ=0\Leftrightarrow T^* =  \frac{T_{ext}}{1+ S^\prime_c/c}.\label{SteadyStateTemperature}
\end{align}

Now as we know the steady state temperature $T^*$ of the information storage system, we still have to describe the dynamics between the original $0$ state and the mutated $1$ state of the single binary digit. It seems reasonable to assume that the molecular structure of the information carrier is not a priori biased for or against any of both possible states, so we assign the same forward and backward temperature dependent reaction rate $kT^*$. $k$ is a generic transition rate between the $0$ and $1$ state. The probability to be in state $0$ then evolves according to

\begin{align}
\frac{dp}{dt} = kT^*(1-2p),
\end{align}

which has a globally stable steady state at $p^*=0.5$. Let us further examine the thermodynamics of this system. The entropy has to be calculated for both states $0$ and $1$:

\begin{align}
S(p) = \ln (p) + \ln (1-p) = \ln (p-p^2).
\end{align}

The time evolution of this entropy is

\begin{align}
S^\prime=\frac{dS}{dt}=\frac{dS}{dp}\frac{dp}{dt}=kT^*\frac{(1-2p)^2}{p-p^2}.\label{EntropyProduction}
\end{align}

From the enumerator $(1-2p)^2$ we see, that $S^\prime=0$ at $p^*=0.5$. $S^\prime>0$ everywhere else except $p=0$ and $p=1$, where $S^\prime\to +\infty$. Thus, the entropy is a Lyapunov function on $p$, which shows a Lyapunov global stability of $p^*=0.5$ as well as $S_*=S(p^*)=\ln 0.25$. This corresponds to the second thermodynamical law of approach to equilibrium, or equivalently, maximum entropy of the system, when the system is not allowed to export its entropy. The $S^\prime\to +\infty$ behavior at $p=0$ and $p=1$ indicates that an infinite export of entropy from the system would be required to sustain these states. These states also correspond to a perfectly determined value of the single binary digit, thus completely error-free storage is never possible.

To maintain any useful information a probability $p>p^*=0.5$ is required. To provide this $p>p^*$ the storage system must be kept away from the $p=p^*$ equilibrium by a constant production of entropy $S^\prime_s$. (The index $s$ stands for suppression of mutations.) The entropy export $S^\prime_c$ is achieved by utilization of metabolic resources. For constant $S^\prime_c$ a non-equilibrium steady state (NESS) is attained. The NESS $p$ value can be calculated from insertion of the actual entropy export for mutation suppression $S^\prime_s$ into (\ref{EntropyProduction}):

\begin{align}
p=\frac{1}{2}\pm\sqrt{\frac{1}{4}-\frac{1}{4+K}},\label{pkEquation}
\end{align}

where $K=S^\prime_s/kT^*$. For $\pm$ we only consider the $+$ alternative, which increases the probability of finding the original value of the digit. The $-$ case would describe the case of an information storage system that utilizes metabolic resources to ensure mutation of digits. We will not bother to discuss this further in this work.

Finally, let us assume that overall metabolic resources $S^\prime_o$ for a single digit are available at a limited rate and can be allocated to cooling the information storage system by $S^\prime_c$ and suppressing mutations by $S^\prime_s$:

\begin{align}
S^\prime_o=S^\prime_c+S^\prime_s.\label{Allocation}
\end{align}

When we optimize this allocation for a maximal $p$ value we can delineate the minimally possible per digit mutation rate for a given $S^\prime_o$. From (\ref{pkEquation}) we see that a maximal $p$ is attained when a maximal $K$ value is taken. We use the above relation (\ref{Allocation}) together with the definition $K=S^\prime_s/kT^*$ and (\ref{SteadyStateTemperature}) and optimize $S_c^\prime\in(0,S_o^\prime)$. We find the optimal $S_c^\prime=(S^\prime_o-c)/2$, which gives the maximal $K$:

\begin{align}
K=\frac{1}{ckT_{ext}}\left(
\frac{S^\prime_o+c}{2}
\right)^2.\label{Kequation}
\end{align}

Substitution into (\ref{pkEquation}) yields an explicit expression for the optimal $p$ given a certain utilization of metabolic resource $S^\prime_o$

\begin{align}
\boxed{p=\frac{1}{2}\left[1+\sqrt{\frac{(1+\overline{s})^2}{C+(1+\overline{s})^2}}\right]}\,,
\end{align}

where $\overline{s}=S^\prime_o/c$ and $C=16kT_{ext}/c$.

We insert the probability of a single binary digit not to be changed from its original value, i.e. not mutated, into an expression for the probability of $m$ binary digits in a sequence of $N$ to be mutated from their original value:

\begin{align}
P^N_m = \left(
\begin{array}{c}
N\\
m
\end{array}
\right)
\left[p(S^\prime)\right]^{N-m}\left[1-p(S^\prime)\right]^m\label{PofM}
\end{align}

\begin{figure*}
\includegraphics[width=\textwidth]{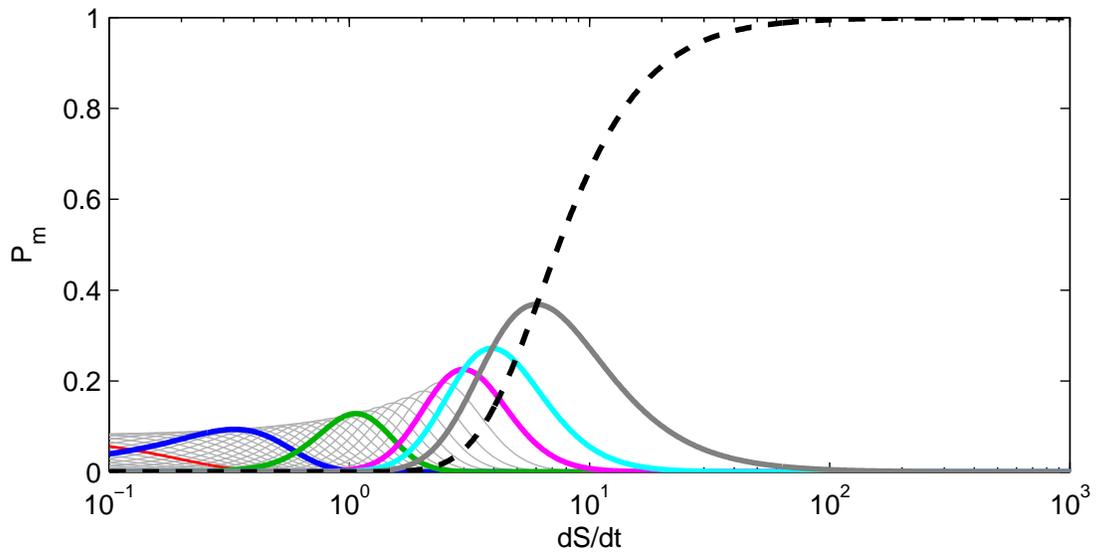}
\caption{{\bf Probability of $m$ mutations for $dS/dt$ metabolic resource utilized on mutation suppression:} Probability $P_m$ of $m$ mutations in a binary sequence vs. entropy production $S^\prime_o=\frac{dS}{dt}$ for mutation suppression. The graph is an evaluation of (\ref{PofM}) for sequence length $N=200$ with $C=200$. The dashed black line represents no mutation $m=0$. The heavy solid lines (Displayed in the following colors from right to left: gray, cyan, magenta, green, blue and red) represent the probabilities $P_m$ of $m=1,2,3,10,20$ and $30$ mutations, respectively.}
\label{MutationProbabilities}
\end{figure*}

This expression describes the probability of $m$ mutations to occur in a sequence of $N$ binary digits given our specific optimal assumptions. Fig. \ref{MutationProbabilities} shows an example for different $S^\prime_0$ values. 

\section*{Stress-induced hypermutation in an adaptation toy model}

\subsection*{General approach}
In the last section we found a dependence of the mutation rate on the level of mutagenic influence and the availability and ability to make use of metabolic resources for mutation suppression. Now we need to understand what the consequences are on adaptation processes. We will assume no recombination, stasis, apoptosis etc. to keep the analysis straight forward. In reality these assumptions would be fulfilled by the simplest unicellular organisms.

First, let us define how stress level and mutations in genetic information storage are connected:

\textit{An elevated stress level a) reduces an organism's ability to deflect mutagenic influence and/or b) restricts or redistributes metabolic resources available to mutation suppression in an unstressed situation and/or c) impairs the means to utilize these resources. Vice versa, any influence exerting one or several of the aforementioned effects can be understood as stress.}

\subsection*{Adaptation toy model}
We are now equipped to construct a generic model of non-recombinatory, mutational adaptation to a rapid change of living conditions. After constructing the general model we will execute and evaluate a numerical simulation. For a population of $N=2^{N_b}$ different binary genotypes $g$ of sequence length $N_b$. We randomly assign $(dS/dt)_g=S'_g$ levels to each genotype $g$. $S'_g$ is the rate of entropy production from utilization of metabolic resource. We introduced before the assumption that $S'$ is set by the stress an organism experiences under certain living conditions. Thus $S'_g$ can also be understood as a stress landscape. To investigate two different adaptation scenarios we assign two types of stress landscape: $m$ mild and $r$ rough. We draw the values in two ways:

\begin{align}
\left(S'_g\right)_{m} = 2D\cdot z_g\mbox{ and }\left(S'_g\right)_r = -D\ln z_g,
\end{align}

where $z_g$ is drawn from a uniform distribution $P(0\leq z \leq 1)=1$. This way both stress landscapes are distributed around the same mean value $D$. The mild one has a flatter distribution of $S'_g$. The rough one has a few high peaks, and mostly low $S'_g$. The mild case should model a gradual adaptation to a less harsh stress type. The rough case should model a sudden drastic change in living conditions with only few beneficial mutations having a strong adaptation effect.

We avoid the commonly used fitness landscape because it is a logical short circuit with respect to our introduced assumption: Fitness is defined as the relative ability of an organism to have its genetic information reproduced by the next generation. As we are trying to understand the effect of mutation rate differences on this very fitness, we should not define a reverse dependency of mutation rate on fitness. It seems conceptually easier to use the effective utilization of metabolic resource $S'$ for mutation suppression as a measure of the stress an organism is facing under a particular set of living conditions.

Each genotype has a certain occupation number $n_g(t)$. This number can increase by inflow of all genotypes $g\in\{1\dots ,N\}$, including $g$ itself. We use a set of linear ordinary differential equations to model the temporal dynamics of the genotype occupation numbers:

\begin{align}
\frac{dn_g}{dt} = \sum_{k}\left[
G(S^\prime_k)P^N_{d(g,k)}(S^\prime_k)
\right].
\end{align}

$d(g,k)$ is the Hamming distance between binary genotypes $g$ and $k$. The Hamming distance is the number of binary digits that are different between $g$ and $k$. This is also the number of mutations by which $g$ and $k$ differ. $G$ is the growth rate of genotype $g$, which we assume to be

\begin{align}
G(S^\prime) = 1-e^{-S^\prime}.
\end{align}

We simulate and compare the simulation results for two types of populations: (1) A population with $S^\prime$-dependent mutation rates, whose dynamics are described above (2) A ``conventional'' population, in which only the growth rates depend on $S^\prime$, the mutation term is simply evaluated for the maximum $\max (S^\prime_g)$ of the whole stress landscape, no matter what the genotypes actual $s^\prime_g$ is. 

We evaluate this toy model for $N_g=5$ in a numerical simulation. The simulation is initialized with all $n_g(t=0)=0$, only the one with the lowest $S_g^\prime$ is assigned $1$. This corresponds to a heavily destabilized wild type. Death and recombination effects are not considered. The dynamics were integrated with a simple first order Euler scheme, step size $0.1$.

The adaptation dynamics of population (1) and (2) are evaluated based on the mean $\overline{S^\prime}=<S^\prime>_g$ across each population. Both populations are initiated and simulated with the same stress landscape $S^\prime_g$. We define a threshold for  $\overline{S^\prime}$, which is $\overline{S^\prime}$ when the population (1) has increased half way from its initial $t=0$ value to its final value. For both populations we determine from the simulation results the times $T^m_{1/2}$ and $T^c_{1/2}$t (for population (1) and (2), respectively), when this threshold is crossed. From these we can calculate a comparative value of the adaptation speed

\begin{align}
\delta T_{1/2} = \frac{T_{1/2}^m-T_{1/2}^c}{T_{1/2}^m}.
\end{align}

$\delta T_{1/2}>0$ indicates faster adaptation of population (1), and $\delta T_{1/2}<0$ indicates faster adaptation of population (2).

\begin{figure*}
\includegraphics[width=\textwidth]{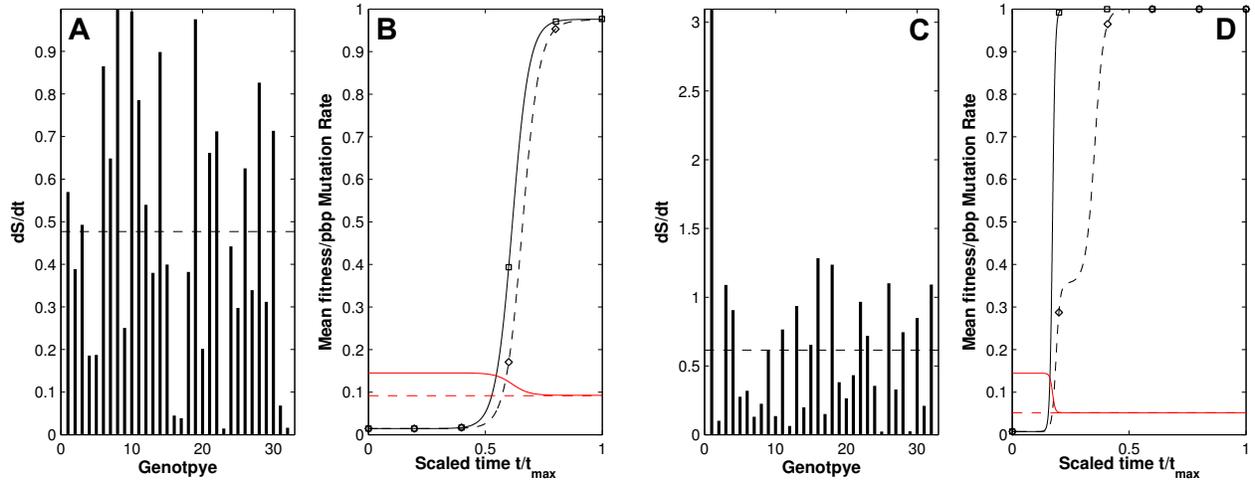}
\caption{{\bf Example stress landscapes and resulting simulation traces:} \textbf{A)} Example stress landscape of mild type \textit{m} \textbf{B)} Example time course of adaptation for stress-induced hypermutation population (1) and conservative population (2) for mild type \textit{m} stress landscape shown in A) \textbf{C)} Example stress landscape of rough type \textit{r} \textbf{D)} Example time courses of adaptation for stress-induced hypermutation population (1) and conservative population (2) for rough type \textit{r} stress landscape shown in C). For each stress landscape type (mild and rough) a stress landscape $(\frac{dS}{dt})_g$ was drawn around a mean of $D=0.5$. Colors, line styles and symbols: The $\frac{dS}{dt}$ landscapes are shown as black vertical bars. In the same graph the mean of these bars is drawn as the horizontal dashed blue line. In the adaptation time courses, the solid (square symbols added for clarity in monochrome print) and the dashed black line (diamond symbols added for clarity in monochrome print) in the right graphs represent the mean fitness $\overline{S^\prime}$ for population (1) and (2), respectively. The solid and the dashed red line (both without symbols for clarity in monochrome print) represent the mean per base pair mutation rate for population (1) and (2), respectively. Parameters used: $D=0.5$, $C=1$, $N=5$, Euler time step $0.1$, simulation executed for 10000 time steps.}
\label{ExampleTraces}
\end{figure*}

\section*{Results}

\subsection*{Mutation rate depends on availability/utilization of metabolic resources}
As can be seen in Fig. \ref{MutationProbabilities}, a decrease in the rate of entropy production by utilization of metabolic resources for protection of the hereditary information $dS/dt$ leads to an increased probability to find higher numbers of mutations in the genetic sequence. $dS/dt$ can be decreased for several reasons: The metabolic resource is not available, the mechanisms to utilize the metabolic resource can be impaired or the metabolic resource is taken up by other organismic processes. While the reasons for a lowered $dS/dt$ can differ, the effects on the mutation number probabilities are the same. 

\subsection*{Hypermutation under stress affects adaptation dynamics}
We already found that the minimally possible mutation rate increases when the metabolic resources utilized for the safekeeping of genetic information are limited. In our toy model for population adaption we can see the effects on adaptation dynamics by the comparison of two populations: Population (1): Both the growth rate $G$ and the mutation rate $(1-p)$ are affected by the level of metabolic resources utilized, Population (2): Only the growth rate $G$ is affected by the level of metabolic resources metabolized.

\textbf{Hypermutation effects in population comparison:}
In Fig. \ref{ExampleTraces} we see how stress-dependence of mutation rates influences adaptation dynamics in comparison to a conservative population with only growth rate depending on metabolic resource utilization. The first, most obvious effect is that on the speed of increase of the mean utilized metabolic resources $dS/dt$ of both populations: In the example traces the stress-dependent mutation rate population adapts visibly faster than the conservative one with only growth rate differences. On a closer look, we can see the driving force – a transient increase of mean mutation rate in the population with stress-dependent mutation rates. The pattern can be observed in many example traces, in the rough as well as mild type stress landscape scenario.

\begin{figure*}
\includegraphics[width=\textwidth]{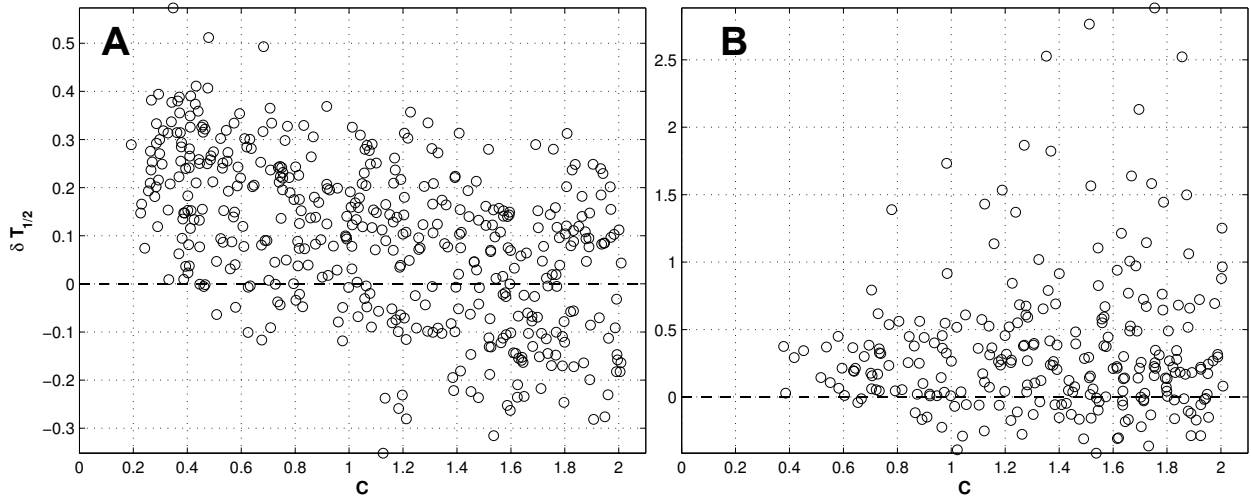}
\caption{{\bf Comparison of adaptation dynamics for different $C$ values:} \textbf{A)} Mild type \textit{m} stress landscape simulation. \textbf{B)} Rough type \textit{r} stress landscape simulation. For each simulation run, the normalized half-time difference $\delta T_{1/2}$ has been plotted vs. the sampled $C$ value. $\delta T_{1/2}>0$ indicates a faster adaptation of the stress-induced hypermutation population (1), $\delta T_{1/2}<0$ indicates a faster adaptation of the conservative population (2). For each condition (mild and rough) 500 simulation runs were executed. Parameters used: $D=0.5$, $N=5$, Euler time step $0.1$, 500 simulations executed for 20000 time steps each, only simulation runs with valid result shown.}
\label{1DComparison}
\end{figure*}

\textbf{$C$ dependence of comparative adaptation dynamics:}
First, we compare for both stress landscape types (mild and rough) the adaptation dynamics for different values of the parameter $C$, which characterizes the mutagenic influence from the environment. In Fig. \ref{1DComparison} we can see the effect of stress-dependent mutation rate increase. For the mild stress landscape, both populations (1) as well as (2) can be faster to adapt, dependent on the structure of the stress landscape. In the rough type stress landscape, faster adaptation of population (1) dominates, while for a some stress landscapes population (2) adapts faster than (1). For the rough type stress landscape extreme acceleration effects are visible. For both stress landscape types the differences in adaptation time course between both populations exhibit a strong dependence on parameter $C$. In the rough type stress landscape high $C$-values are associated with faster adaptation of population (1). In the mild type stress landscape population (2) mostly adapts faster at high $C$ values.

\begin{figure*}
\includegraphics[width=\textwidth]{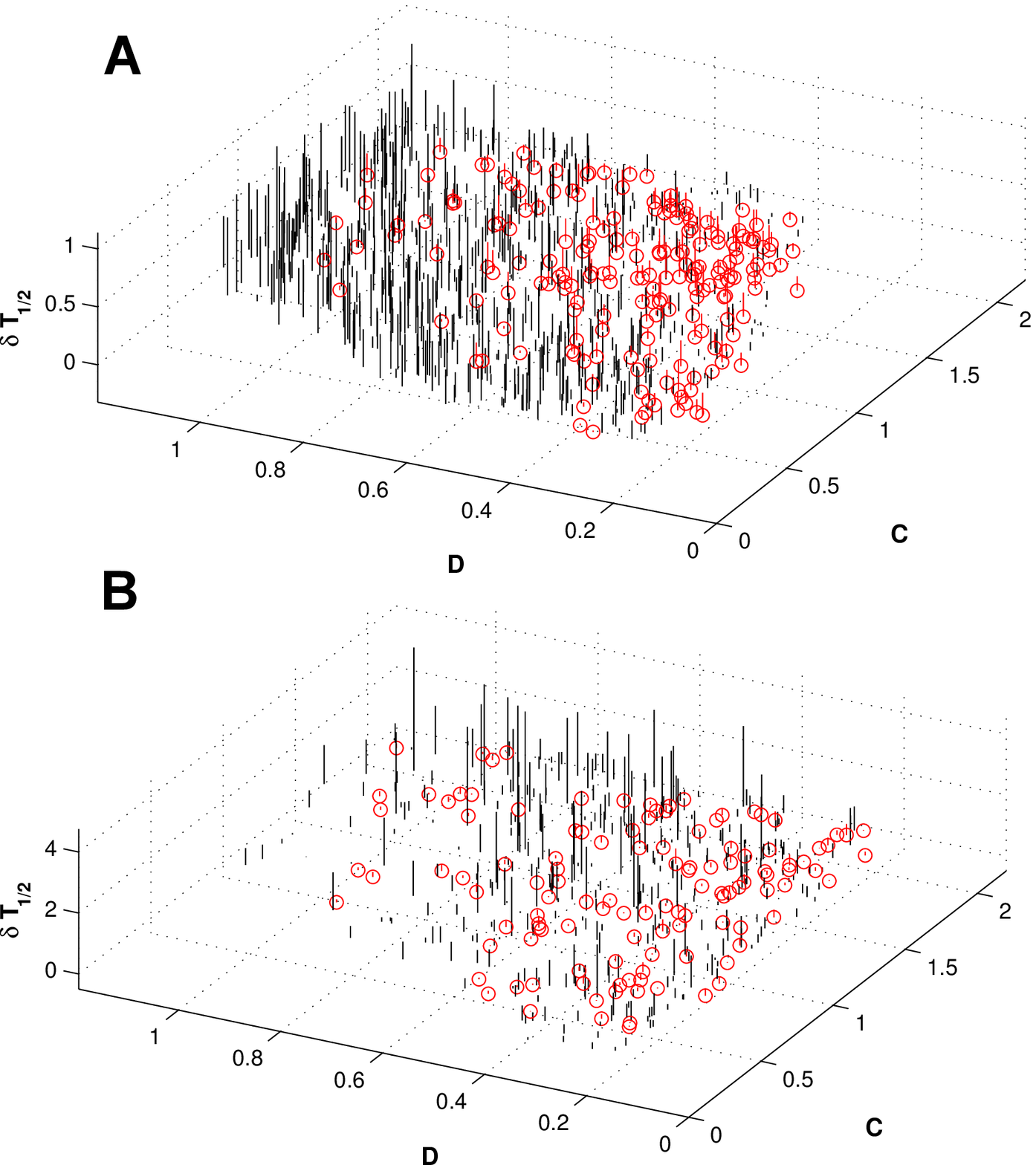}
\caption{{\bf Comparison of adaptation dynamics for $(C,D)$ sample points:} \textbf{A)} Mild type \textit{m} stress landscape simulation. \textbf{B)} Rough type \textit{r} stress landscape simulation. For each stress landscape type (mild and rough) 1000 simulations have been executed. All those that reached sufficient growth during the simulation and for which $|\delta T_{1/2}|>0.01$ have been plotted into the above diagrams. Black lines represent examples in which the stress-induced hypermutation population (1) adapted faster, i.e. $\delta T_{1/2}>0$. Red lines with a red circle represent a faster adaptation of the conservative population (2) adapting faster, i.e. $\delta T_{1/2}<0$. Parameters used: $N=5$, Euler time step $0.1$, 1000 simulations executed for 15000 time steps, only simulation runs with valid results are shown. In the big regions without any sample points the simulation could not long enough to be properly evaluated.}
\label{2DComparison}
\end{figure*}

\textbf{$(C,D)$ dependence of comparative adaptation dynamics:}
Second, we compare for both stress landscape types (mild and rough) the adaptation dynamics for different values of the parameter duplet (C,D). In Fig. \ref{2DComparison} different typical regions show up:
In the mild type stress landscape, the samples describe a kind of ``wedge'' that expands from $(C=0.2,D=0.2)$ and grows in $D>0$ direction for increasing $C$. Inside this wedge the (2) population exhibits faster adaptation dynamics. In the rest of the parameter space accelerated adaptation of the (1) population dominates.
In the rough type stress landscape case no clear regions can be identified, but mostly adaptation acceleration of population (1) is more frequent and the acceleration effect is stronger. Sample (C,D) points showing faster acceleration of population (2) exhibit a trend to decrease in occurrence in direction of increased $D$. Especially in the rough type stress landscape, examples of “hyper-adaptability” occur, where up to 4-fold relative acceleration of population (1) arise from the stress-dependence of mutation rate.

\section*{Discussion}

The logic of the presented work is quite short and simple:

\textit{A higher rate of primary mutations as well as a lowered ability to employ metabolic resources in mutation suppression increase the minimum effective mutation rate. This predicts transient mutation rate increases as a response to stress, which can accelerate adaptation.}

The logic is well-described by these two sentences, and the thermodynamical and quantitative analysis are just a more formal way of deriving and consolidating this logic. Still, while the above logic is common fare amongst many biologists and medical researchers, the very specialists concerned with the topic have not actually reached a final common sense. Much experimental evidence indicates that the above logic applies \cite{GordoScience, StressMutationsEColi, AmplificationMutationSeparate, StressDependenceYeast, StressDependenceChlamydomonas, StressDependenceDrosophilaOne, StressDependenceDrosophilaTwo} and its consequences are discussed \cite{EffectsOfStressDependence,Selection}. However, we have not come across a derivation of above logic from fundamental physical axioms. Our work might add fundamental physical credibility to the aforementioned experimental evidence of stress-induced hypermutation and the resulting possibility of acceleration of adaption by transient mutation rate increase.

We developed a purely thermodynamical, non-mechanistic analysis, which can by definition point out specific mechanisms. Instead, it can delineate the physical limitations, which biological information storage systems can not exceed. Experimental researchers find a whole range of mutagenic genome repair mechanisms and their resultant genetic mutation rates\cite{DNARepairReview,SOSResponseReview,MutationStressResponse}. A concept like the one we presented herein could possibly help to order some of this abundance of experimental knowledge.

\textbf{Specific selective scenario vs. physical constraint:}
Hypermutation under stress and accelerated adaptation by transient mutation rate increase (Adaptive Mutation) demand an evolutionary explanation.

First of all, it seems perfectly credible to use frequently changing living conditions of simple single cell organisms as a selective environment that favors adaptive agility. However, the explanation we suggest in this work is a physical constraint all cells are subject to, which does not require a specific selection scenario. It does not seem unreasonable, that mechanisms of genome repair and associated mutation suppression have developed to get close to the thermodynamical optimum. Both explanations, a specific selection scenario as well as the described physical constraint, lead to similar evolutionary development. However, latter explanation by a physical constraint needs weaker assumptions and applies more generally than one based on a specific selection scenario. Let us spell out the conditions which my analysis is based on: (1) The thermodynamical laws must hold, (2) an organism utilizes metabolic resources to maintain the integrity of its stored genetic information and (3) under stress this capability is compromised. This should most likely apply to all organisms and consequently the pattern of stress-induced hypermutation should be observable in many if not all of them. Note that a physical constraint and a specific selective scenario are in no way mutually exclusive.

\textbf{Discussion of findings in relation to other responses to compromised genetic integrity:}
A physical constraint as a reason for stress-induced hypermutation would be effective in all organisms. However, especially higher multi-cellular organisms can be expected to have developed counter-strategies to keep free of mutation effects in their constituent cellular organisms. Apoptosis or cellular senescence should be mentioned here.

Gene repair and proofreading pathways are studied mostly in the context of DNA replication, see \cite{DNARepairReview,SOSResponseReview}. Differences in DNA replication also have pronounced effects on adaptation and most likely these effects and those from stress-induced hypermutation both influence adaptation processes.

\textbf{Discussion of findings in relation to other studies:} We are aware of only two other theoretical studies that investigate the effects of stress-induced hypermutation on adaptation. \cite{EffectsOfStressDependence} shows that in stable, steady state populations stress-dependence increases genetic loads, as well as the cost of the maintenance of sex. \cite{ConditionDependenceSex} investigates the influence an antiproportional relation between mutation rate and ornamentation has on sexual selection behavior and its effectiveness to avoid mates with germ-line mutations. In contrast, our work explores the physical credibility of stress-induced hypermutation as well its role in the time course of adaptation, which should prove especially interesting in medical scenarios of rapid pathogen evolution.

While not directly concerned with the topic of this work, an abundance of theoretical literature exists on the role of mutation rates in evolution. Generally, mutation rates are assumed to be constant across genotypes. Some works do allow for mutation rate differences between genotypes\cite{EigenRundumschlag, EarlyReplicons}. Most interestingly, it has already been recognized, that populations can exhibit accelerated adaptation by a transient increase of mutator alleles, which exhibit the genetic trait of higher mutation rates\cite{MutatorAllelesAdaptation}. In contrast, a transient increase in mutation rate induced by stress is not a genetic trait, but simply an organism's incapability to sustain safe storage of genetic information under stress conditions. Except \cite{EffectsOfStressDependence, ConditionDependenceSex} and our own approach presented herein. We know of no other theoretical work to investigate the role of such transient, stress-induced increases of mutation rates in adaptation.

Also, the question if mutation rates are a selectable, genetic feature has been discussed at large, e.g see \cite{EvolutionOfMutationRatesReview, EvolutionOfMutationRatesExperimental, EvolutionOfMutationRates, Oenococcus, MutationRateSelfDependence}. \cite{MutationRateVariationMulticellular} gives a comprehensive overview how living conditions of an organism shape its mutation rate as a genetic trait. The bigger picture of recent studies on mutation rates, however, challenges the whole notion of a typical, mutation rate for a certain genotype, and instead suggests condition-dependent, dynamically changing mutation rates \cite{AntibioticsVariabilityReview, MutationStressResponse}. Our work might contribute fundamental credibility and insight to this notion of dynamically changing, condition-dependent mutation rates.

A further related topic is that of the evolution of evolvability\cite{EvolutionOfEvolvability,RobustnessEvolvability}, and especially the causality between variation and selection is a matter of concern.\cite{EvolutionOfEvolvabilityTwo} As stated earlier, stress-induced hypermutation could indeed be a case of evolved evolvability, but the physical constraint we present in this work seems to be of more general applicability. The introduction of \cite{EvolutionOfEvolvabilityTwo} gives an overview of experimental results, theoretical work and discussion related to the evolution of evolvability. With regard to causality, the interdependence of stress level and variation as per appearance of new mutants is reminiscent of a closed regulatory loop: A higher stress level induces a higher mutation rate, which allows swift adaptive mutation, which in turn relieves stress, and finally lowers the mutation rate again. Note, that this regulatory circle arises from a fundamental physical constraint, not as the result of a specific selective scenario.

Another general class of works is concerned with the application of statistical mechanics methods to the study of the evolution on the DNA and RNA sequence level, e.g. see \cite{DarwinianToThermodynamics, SelfOrganizationNaturalSelection, MaximumPrinciple, MethodStuff}. Much of this work is of methodological value, and rests on assumptions, which are challenged and rectified by works of the type of those mentioned shortly before.

\textbf{Experimental avenues:}
(1) Experimental studies of the role of mutation rates in adaptation face a big conceptual problem: Mutations take effect alongside with other selective forces, e.g. growth and death rate differences. Our work suggest, that cells kept in non-growth but viable conditions still amass mutations in their DNA. Further, the probability and frequency of mutations should depend on the level of additionally imposed stress. Thus, an experimenter could keep a population of cells on non-growth medium, while applying differential stress levels associated with different mutation probabilities and frequencies. Migration back to normal medium allows assessment of only the inflicted mutations, separate of other adaptation effects.
(2) A higher rate of primary mutations as well as a lowered ability to employ metabolic resources in mutation suppression can be caused by various stress types, the induction of hypermutation should be a common outcome for all of them. Thus, in the aforementioned experiment, the exact type of stress should be irrelevant to the observation of hypermutation. Further, different stress types should actually jointly increase the hypermutation response, which suggests experiments with combined stress exposure.

\textbf{Medical implications:}
On an evolutionary scale, mutations are the source of development and biodiversity. On the scale of the human individual, however, mutations are mostly associated with undesirable, often even critical consequences. A possible cause of cancer are genetic mutations and the results presented in our work could indicate that increased stress in general increases genetic mutation rates and thereby the risk of cancer. Cancer resistances to chemotherapy as are supposedly driven by mutations. Our results indicate that chemotherapy could in fact induce hypermutation in cancer cells and thereby amplify their mutational adaptation. Similarly, our results support the view that antimicrobials do not only attack a pathogen but at sublethal doses also increase its adaptability by the stress inflicted upon the pathogen.\cite{AntibioticsVariabilityReview}

All mentioned phenomena are characterized by surprisingly swift adaptability of the respective cells, and the results from this work point towards stress-induced hypermutation as a possible common reason for this rapid adaptation. Firstly, this can deepen our understanding of the underlying causality, and secondly indicates, that therapeutic approaches used to disrupt the adaptability in one scenario, could possibly be adapted to the other scenarios.

\section*{Supplementary Materials}
MatLab scripts to produce the displayed simulation results.

\section*{Acknowledgments}
The first idea was triggered by a discussion with Shilan Mistry. Discussions with the following were essential: Siddharth Arora, Baboo Sabyasachi, David Hasselbach, Michael C. Mackey, Moises Santillan, Bartek Borek, Gregor Fussmann, Luke McNally, Peter Richter and Isabel Gordo. All anonymous reviewers' efforts are deeply appreciated.\\
Financial support from NSERC (Natural Sciences and Engineering Research Council of Canada), MITACS (Mathematics of Information Technology and Complex Systems) and a student and a traveling fellowship of the Studienstiftung des Deutschen Volkes. No funding body influenced the conception, execution or writing of this work.

\end{document}